\documentclass[preprint2]{aastex61}

\usepackage{graphicx}
\usepackage[suffix=]{epstopdf}
\usepackage{natbib}
\usepackage{amsmath}
\usepackage{url}
\usepackage{xspace}

\newcommand{\Kepler}{\textsl{Kepler}\xspace}

\begin{document}

\title{The GALEX View of ``Boyajian's Star'' (KIC 8462852)}

\shorttitle{GALEX View of ``Boyajian's Star''}
\shortauthors{Davenport et al.}

\author{James. R. A. Davenport}
\affiliation{Department of Physics \& Astronomy, Western Washington University, 516 High St., Bellingham, WA 98225, USA}
\affiliation{NSF Astronomy and Astrophysics Postdoctoral Fellow}

\author{Kevin R. Covey}
\affiliation{Department of Physics \& Astronomy, Western Washington University, 516 High St., Bellingham, WA 98225, USA}

\author{Riley W. Clarke}
\affiliation{Department of Physics \& Astronomy, Western Washington University, 516 High St., Bellingham, WA 98225, USA}

\author{Zachery Laycock}
\affiliation{Department of Physics \& Astronomy, Western Washington University, 516 High St., Bellingham, WA 98225, USA}

\author{Scott W. Fleming}
\affiliation{STScI, 3700 San Martin Dr., Baltimore, MD 21218, USA}

\author{Tabetha S. Boyajian}
\affiliation{Department of Physics and Astronomy, Louisiana State University, 261-A Nicholson Hall, Tower Dr, Baton Rouge, LA 70803, USA}

\author{Benjamin T. Montet}
\affiliation{Department of Astronomy and Astrophysics, University of Chicago, 5640 S. Ellis Ave, Chicago, IL 60637, USA}
\affiliation{NASA Sagan Fellow}

\author{Bernie Shiao}
\affiliation{STScI, 3700 San Martin Dr., Baltimore, MD 21218}

\author{Chase C. Million}
\affiliation{Million Concepts LLC, PO Box 119, 141 Mary St, Lemont, PA 16851, USA}

\author{David J. Wilson}
\affiliation{Department of Physics, University of Warwick, Coventry CV4 7AL, UK}

\author{Manuel Olmedo}
\affiliation{Instituto Nacional de Astrof\'{i}sica Optica y Electr\'{o}nica, Tonantzintla, Puebla, M\'{e}xico}

\author{Eric E. Mamajek}
\affiliation{Jet Propulsion Laboratory, California Institute of Technology, 4800 Oak Grove Drive, Pasadena, CA 91109, USA}
\affiliation{Department of Physics \& Astronomy, University of Rochester, Rochester, NY 14627, USA}

\author{Daniel Olmedo}
\affiliation{Instituto Nacional de Astrof\'{i}sica Optica y Electr\'{o}nica, Tonantzintla, Puebla, M\'{e}xico}

\author{Miguel Ch\'{a}vez}
\affiliation{Instituto Nacional de Astrof\'{i}sica Optica y Electr\'{o}nica, Tonantzintla, Puebla, M\'{e}xico}

\author{Emanuele Bertone}
\affiliation{Instituto Nacional de Astrof\'{i}sica Optica y Electr\'{o}nica, Tonantzintla, Puebla, M\'{e}xico}

\begin{abstract}

The enigmatic star KIC 8462852, informally known as ``Boyajian's Star'',  has exhibited unexplained variability from both short timescale (days) dimming events, and years-long fading in the \Kepler mission. No single physical mechanism has successfully explained these observations to date.
Here we investigate the ultraviolet variability of KIC 8462852 on a range of timescales using data from the GALEX mission that occurred contemporaneously with the \Kepler mission. The wide wavelength baseline between the \Kepler and GALEX data provides a unique constraint on the nature of the variability.
Using 1600 seconds of photon-counting data from four GALEX visits spread over 70 days in 2011, we find no coherent NUV variability in the system on 10--100 second or months timescales.
Comparing the integrated flux from these 2011 visits to the 2012 NUV flux published in the GALEX-CAUSE Kepler survey, we find a 3\% decrease in brightness for KIC 8462852. We find this level of variability is significant, but not necessarily unusual for stars of similar spectral type in the GALEX data.
This decrease coincides with the secular optical fading reported by \citet{montet2016}. We find the multi-wavelength variability is somewhat inconsistent with typical interstellar dust absorption, but instead favors a $R_V=5.0\pm0.9$ reddening law potentially from circumstellar dust.
\end{abstract}

\section{Introduction}
KIC 8462852, also known as ``Boyajian's Star'', is an unusual F3 dwarf in the \Kepler field that has exhibited unexplained optical variability on a variety of timescales. The initial discovery was of several dramatic, short timescale (days) dimming events with amplitudes up to 20\% in the \Kepler 30-min cadence data \citep{boyajian2015}. Though the \Kepler mission \citep{borucki2010} obtained data at a 30-min cadence for $\sim$4 years on this star, no definitive pattern or cycle was found, nor has any single explanation for this variability been accepted by the community \citep{wright2016b}.

Analysis of archival optical photographic plates has found that KIC 8462852 may have additionally faded nearly 16\% over the past century \citep{schaefer2016}. Such a precise measurement for a single star is difficult, and the result has been debated \citep{hippke2016}. However, using the 53 ``Full Frame Images'' (FFIs) spread over the 4-year \Kepler mission, \citet{montet2016} were able to trace the brightness of KIC 8462852 using an independent flux calibration. The resulting flux-calibrated FFI light curve showed definitively that KIC 8462852 faded by more than 3\% over 4 years. A years-long timescale variability, with possible periodicity, has recently been confirmed with an analysis of archival ground-based optical photometry \citep{simon2017}.

The short (days) and long (years) timescale variability discovered for KIC 8462852 has presented a unique set of observational constraints for any single model used to describe the system. For example, if variable dust extinction is responsible for both temporal features, then the dust must have a wildly variable density distribution on small spatial scales, and a small density gradient over large spatial scales. Searches for an infrared flux excess consistent with a foreground or circumstellar dust shell have found no strong detection \citep[e.g.][]{marengo2015}, further complicating attempts to attribute the variability to dust structures.

Since optical variability and infrared follow-up has not produced a robust explanation for KIC 8462852, further multi-wavelength studies are needed to constrain the nature of the long timescale fading and short timescale dimming. Multi-band photometric and spectroscopic campaigns are underway\footnote{\url{http://www.wherestheflux.com}}, which will provide an improved understanding of any future ``dips''. However, no multi-wavelength  measurement of the mysterious variability for KIC 8462852 contemporaneous with the \Kepler observations has been analyzed.

Archival photometry at ultraviolet wavelengths from the GALEX mission \citep{galex}
spanning a range of timescales from seconds to more than a year is now available. This unique set of observations occurred during the \Kepler mission, providing an independent constraint on the variability discovered in the \Kepler photometry for KIC 8462852. The wide wavelength range probed by GALEX and \Kepler also allows us to explore models of the variability based on dust extinction and thermal cooling.

The various GALEX data products used in our analysis are introduced in \S\ref{sec:data}. In \S\ref{sec:short} we analyze the NUV data over 10--100 second timescales. In \S\ref{sec:long} we explore the long timescale evolution of KIC 8462852 between the 2011 and 2012 visits, and compare directly to the observed fading by \citet{montet2016}.
In \S\ref{sec:dust} we discuss possible interpretations for the nature of KIC 8462852 that the combined \Kepler and GALEX observations provide, including an estimate of the dust extinction properties necessary to reproduce the long timescale NUV observations. Finally in \S\ref{sec:summary} we summarize this work, and discuss the potential utility of GALEX in the study of other rare and unusual variable \Kepler objects.

\section{GALEX Observations}
\label{sec:data}

\begin{figure}[!t]
\centering
\includegraphics[width=2.5in]{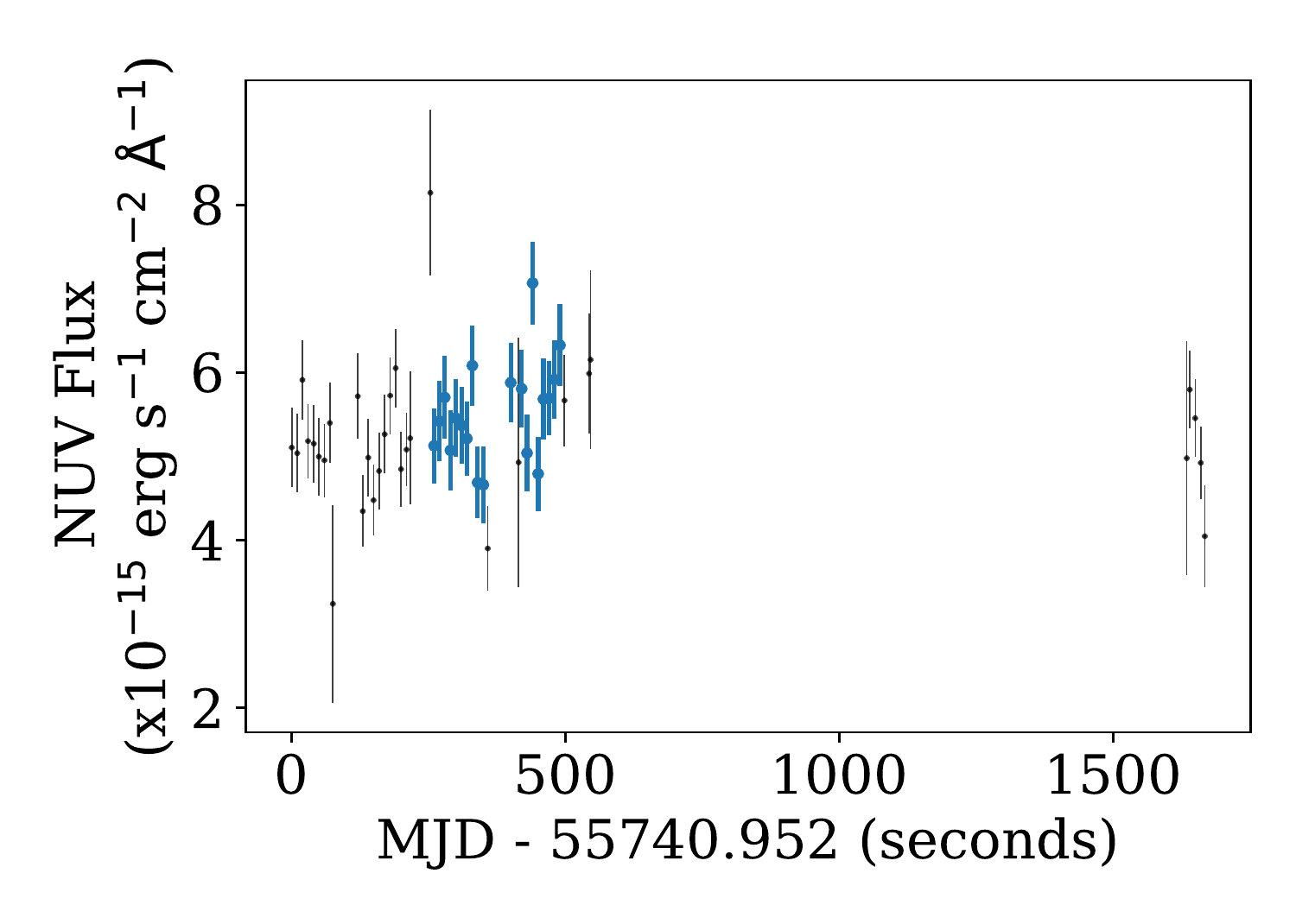}
\includegraphics[width=2.5in]{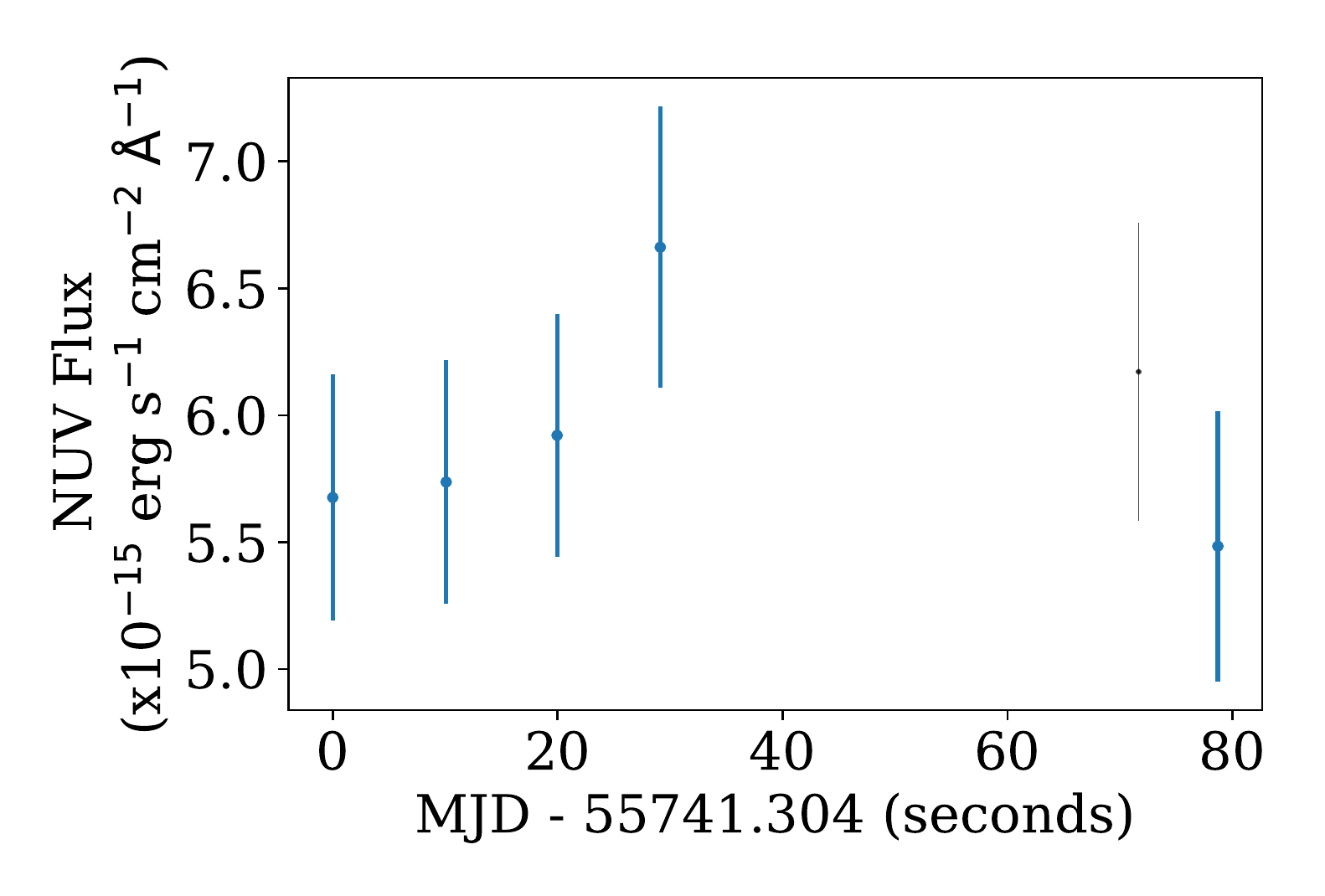}
\includegraphics[width=2.5in]{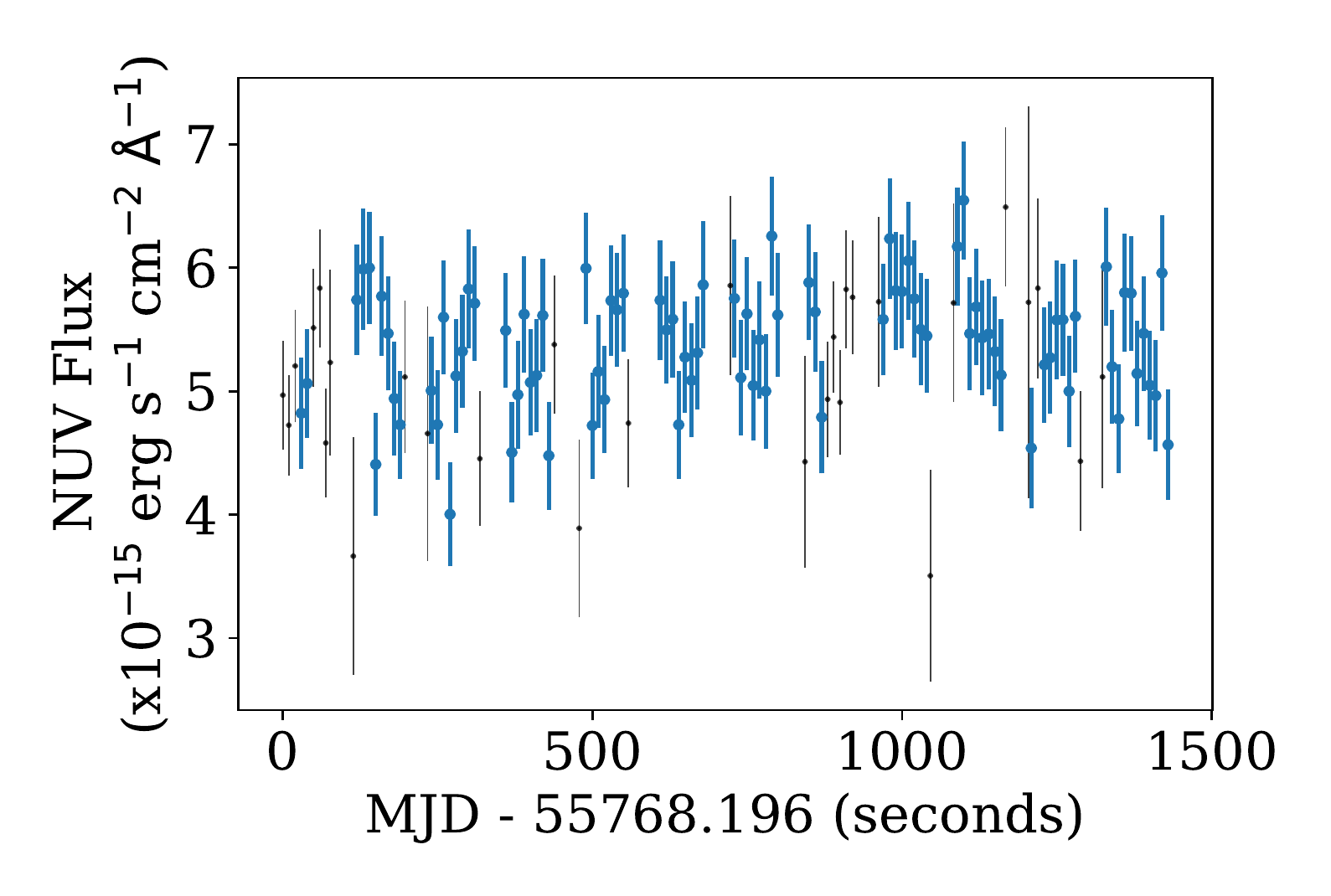}
\includegraphics[width=2.5in]{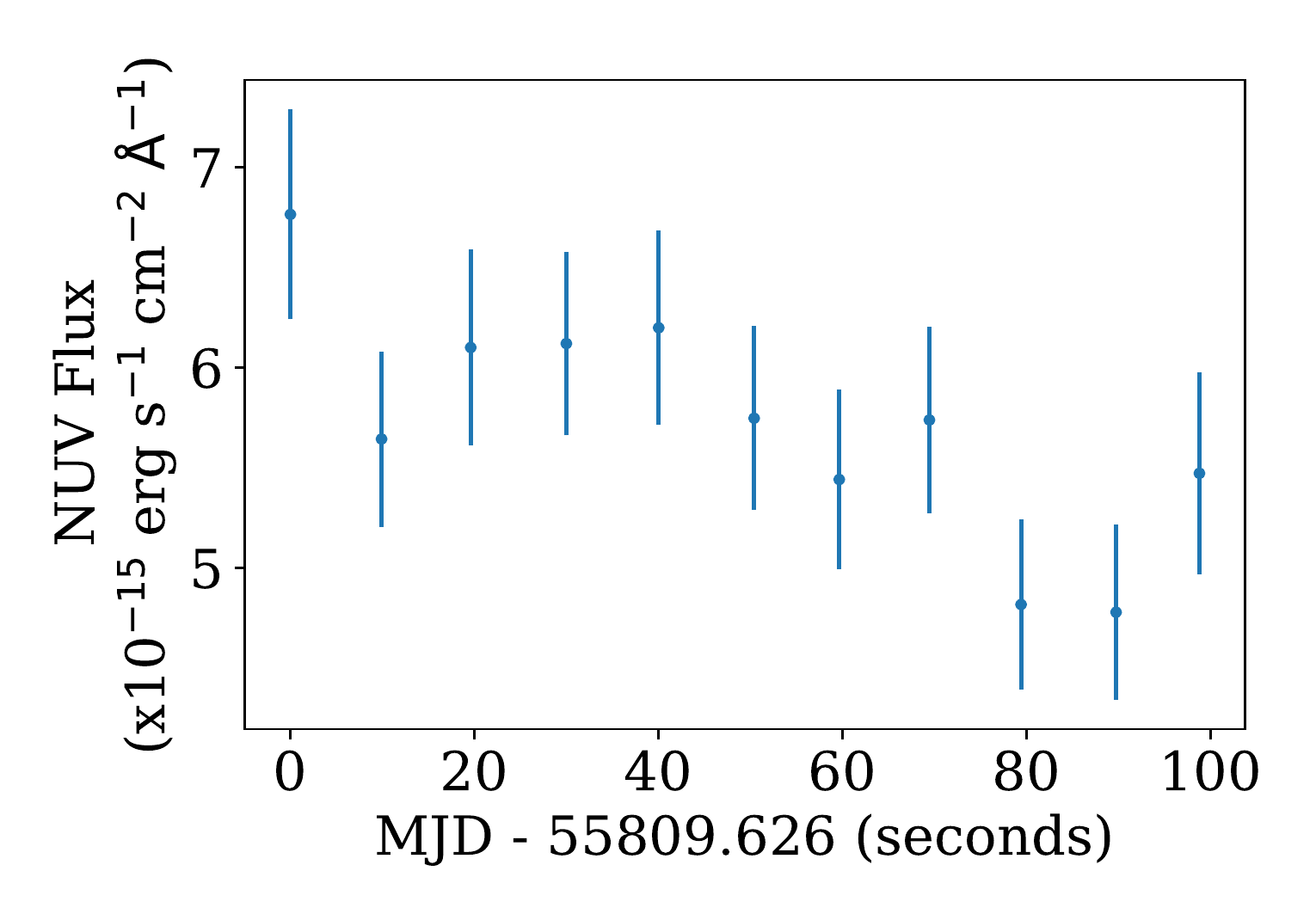}
\caption{
Light curves from {\tt gPhoton} sampled at a 10-second cadence for the 4 visits in 2012. All epochs are shown (grey), while those having no photometric warning flags set are highlighted (blue). Error bars shown are the photometric errors for each point computed by {\tt gPhoton}.}
\label{fig:shorttime}
\end{figure}

Time-tagged photon data has recently become available for GALEX \citep{million2016}, including a Python toolkit to search for and interact with this high cadence data product called {\tt gPhoton} \citep{gphoton}. This allows us to resample the GALEX main survey data into any desired cadence. In the case of KIC 8462852, the primary GALEX survey obtained $\sim$1600 seconds of data during four separate visits spread across a $\sim$70 day baseline in 2011. These high-cadence data from 2011 are also co-added as part of the GALEX ``GR6'' data release \citep{bianchi2014}. In this work we analyze only the NUV data ($\lambda_{eff}=2315.7$\AA), as KIC 8462852 is too faint in the GALEX FUV band. Note the GALEX NUV band is similar in wavelength coverage to the Swift $uvm2$-band analyzed for KIC 8462852 by \citet{meng2017}.

As part of the GALEX Complete All-Sky UV Survey Extension (CAUSE) program, 104 square degrees within the \Kepler field were re-observed in the NUV, creating the GALEX-CAUSE Kepler survey (hereafter GCK).
This survey occurred in 2012, and overlapped a portion of the Quarter 14 operations from the original \Kepler mission. The GCK data was obtained using scan-mode observing that differed from the standard GALEX survey.
A catalog of the integrated fluxes and uncertainties for 475,164 \Kepler targets observed in GCK, including for KIC 8462852, was made available by \citet{olmedo2015}. In the case of KIC 8462852, the GCK catalog utilizes 1413.8 seconds of integration in 2012. Unfortunately, since the observing mode differed from the standard GALEX survey,  GCK data is not available for time-series analysis with {\tt gPhoton} presently.

\section{Short Timescale Variability}
\label{sec:short}

Within each of the four primary mission GALEX visits available for KIC 8462852 we searched for short timescale variability using {\tt gPhoton}\footnote{Using {\tt gPhoton} version 1.28.2}. While nano-second optical variability has been investigated for this target \citep{abeysekara2016}, few other studies have looked at variability on timescales shorter than the 30-minute cadence available with \Kepler.
The four GALEX visits in 2011 ranged from $\sim$70 to $\sim$1400 seconds in duration. Data for each visit was sampled at a 10-second cadence with {\tt gPhoton}, as shown in Figure \ref{fig:shorttime}. Small amplitude variability is apparent in several of the visits, with coherent structure over durations of approximately 60-100 seconds. Computing a Lomb-Scargle periodogram using {\tt gatspy} \citep{gatspy} on the entire {\tt gPhoton} light curve, we find moderate power with a broad peak at around 80-seconds. This appears to be due to the $\sim$120 second observing cycle of the GALEX instrument in the standard ``Petal Pattern'' observing mode, and we believe is not astrophysically significant.

A periodic signal of 0.88 days was also found in the \Kepler photometry, which was presumed by \citet{boyajian2015} to be due to the rotation of starspots in- and out-of view on the surface of KIC 8462852. Each of the four GALEX visits shown in Figure \ref{fig:shorttime} are too short to entirely capture this rotation signature. Our periodogram computed using all four of the {\tt gPhoton} light curves together also does not show any signs of this 0.88 day period.

\begin{figure}[!t]
\centering
\includegraphics[width=3.5in]{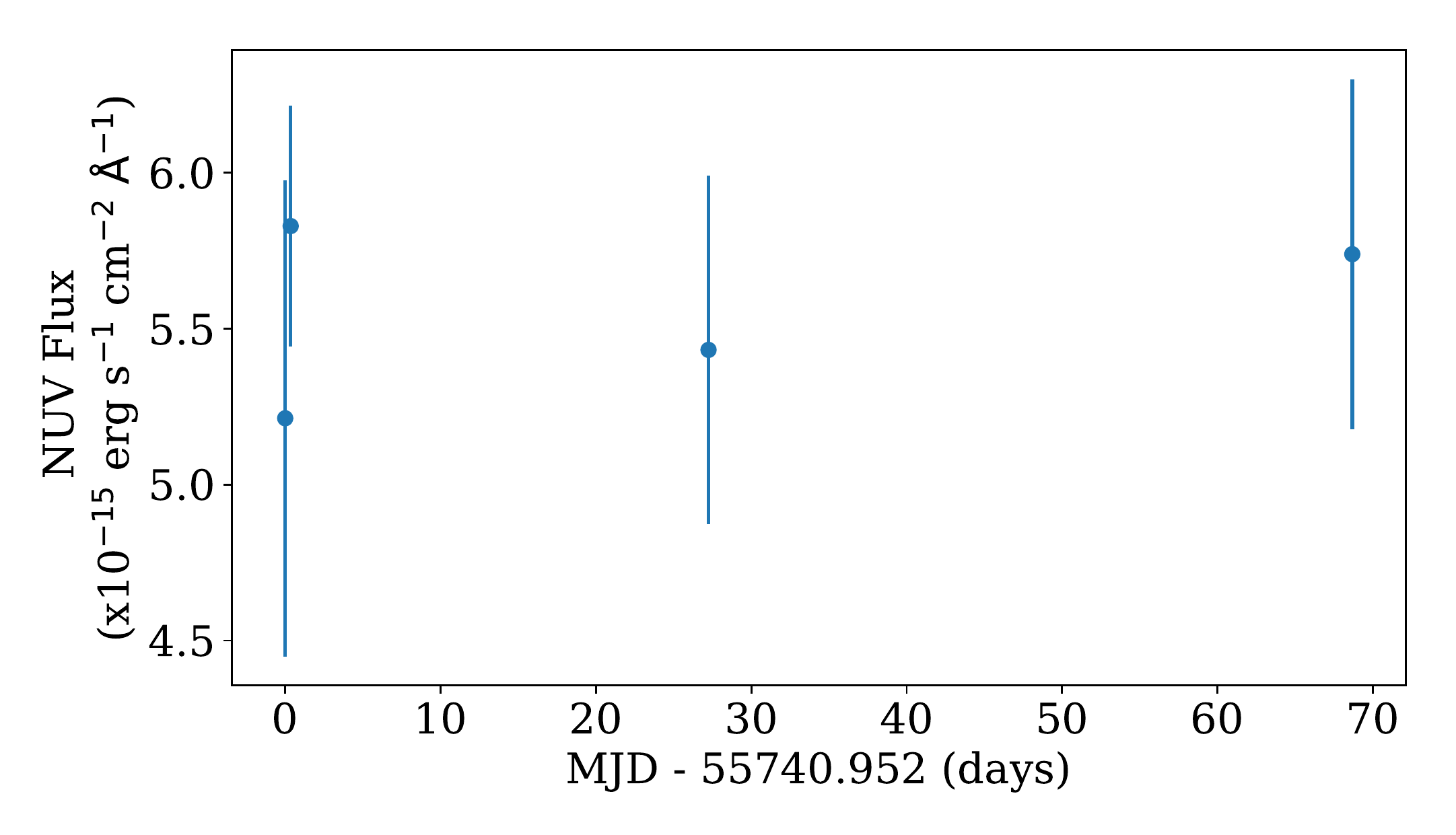}
\caption{Median flux within each of the four visits spaced over $\sim$70 days in 2011 by GALEX. Uncertainties shown are the standard deviation in flux within each 10-sec sampled {\tt gPhoton} light curves from Figure \ref{fig:shorttime}. No significant change in flux is seen over this 70 day window.
}
\label{fig:medtime}
\end{figure}

Since the standard GALEX data for this target was spread over four separate visits, we also examined the medium-timescale variability over $\sim$70 days. In Figure \ref{fig:medtime} we show the median flux from each of the four {\tt gPhoton}-processed visits. The uncertainties shown are computed as the standard deviation in the 10-sec sampled data within each visit, and are $\sim$10x larger than the statistical error on each visit's median flux.  Though there is scatter between these four visits in Figure \ref{fig:medtime}, no significant coherent variability is seen on this intermediate timescale with GALEX. Unfortunately this 70-day time window also did not correspond to any of the previously identified dimming events from \citet{boyajian2015}.

\section{Long Timescale Variability}
\label{sec:long}

\begin{figure}[!t]
\centering
\includegraphics[width=3.5in]{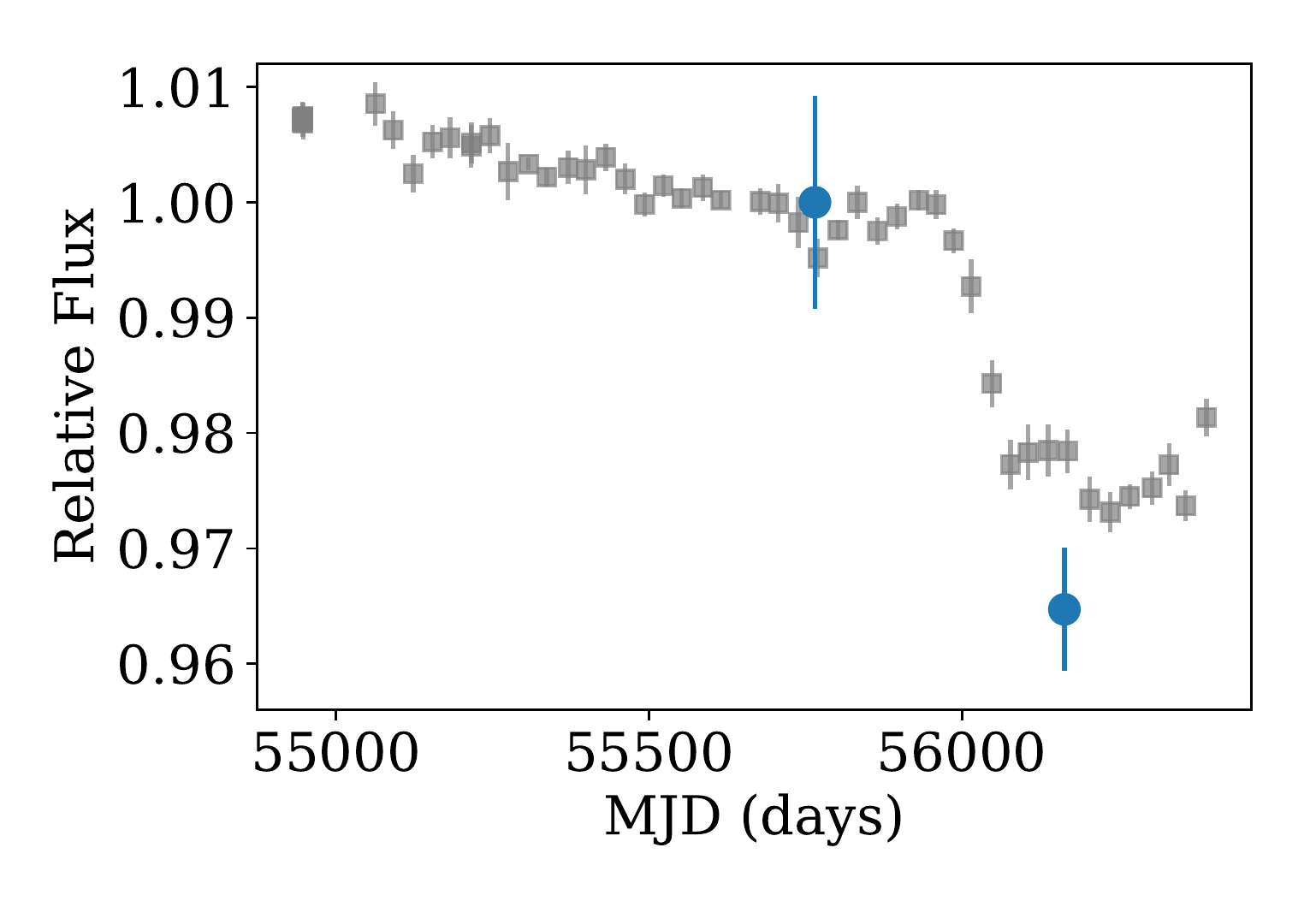}
\caption{
Comparison of the 2011 and 2012 fluxes for KIC 8462852 as measured by GALEX (blue circles), with the \Kepler FFI data shown in \citet{montet2016} as reduced with the new ``f3'' package from \citet{montet2017} for comparison (grey squares). The amplitude of variability over this time window is nearly identical between the two surveys.
}
\label{fig:longtime}
\end{figure}

In Figure \ref{fig:longtime} we present the GALEX data for this target as observed in 2011 and 2012. The 2011 data represents the final GALEX GR6 catalog flux value for KIC 8462852 of $16.46 \pm 0.01$ mag from \citet{bianchi2014}, and is the integrated flux from all four visits described in \S\ref{sec:short}. The 2012 data is from the GCK data from \citet{olmedo2015}, and measured a NUV brightness for KIC 8462852 of $16.499\pm0.006$ mag. Both apparent NUV magnitudes were converted to fluxes, and then  normalized to the flux of the 2011 visit. This results in a measured NUV fading of 3.5 $\pm$ 1.0\%.

In Figure \ref{fig:longtime} we also show the slow fading discovered in the \Kepler FFI's by \citet{montet2016}. Note: the fact that the GALEX and \Kepler FFI data are normalized to a relative flux of 1 around 2011 (MJD$\sim$55700) is a coincidence. However, the observation that the GALEX flux decays coherently with the \Kepler FFI flux over this time baseline is significant.

To determine what the typical variation in NUV flux is for F stars, we analyzed the variability for over 140,000 GCK stars in common with the Kepler Stellar Catalog \citep{mathur2017}, observed on average 15 times over a time interval of 40 days in 2012. Note the original GALEX data was not available over the entire Kepler footprint, and was not taken over a single 70 day observing window in 2011 for all targets as for KIC 8462852. A full analysis of this variability, while beyond the scope of our work here, is underway (Daniel  Olmedo et al. in preparation). We found that stars with temperatures near KIC 8462852 (6750 K) have an average variation of 3.5\%. 
GALEX was calibrated using the white dwarf LDS749b, which was repeatedly observed during normal operations. Figure 6 from \citet{million2016} finds visit-to-visit scatter of the photon-level data for LDS749b of 2-3\% using {\tt gPhoton}. The GALEX calibration work done by \citet{morrissey2007} finds the photometric repeatability for stars at the brightness of KIC 8462852 is $\pm$1.5\%. GALEX also provided an estimate for longer-exposure repeatability as a function of magnitude, which indicates an expected $\sim$3\% uncertainty for our target.
\footnote{\url{https://asd.gsfc.nasa.gov/archive/galex/FAQ/counts_background.html}}
The NUV variation seen for KIC 8462852 is therefore not necessarily abnormal for stars of this spectral type.

As an aside, we also searched for long timescale variability for KIC 8462852 in the infrared from the WISE single-exposure source database using the W1-band (3.4$\mu$m). This dataset from the original WISE mission \citep{wise}, and the NEOWISE extended mission \citep{mainzer2014} provides $\sim$2-day clusters of photometry spaced every 6 months due to the spacecraft roll pattern. Unfortunately the GALEX observations for KIC 8462852 occurred during the observation gap between WISE and NEOWISE, and thus a direct comparison between the NUV and IR is not possible here.
We found no clear  long-term variability spanning 2009 through 2017 for KIC 8462852 in the W1-band. However, a more detailed comparison of this rich IR dataset to the recently published work from \citet{simon2017} and \citet{meng2017} is warranted.

\section{Implications for the Nature of KIC 8462852}
\label{sec:dust}

While many explanations for the nature of KIC 8462852 have been proposed, there is effectively no consensus on the nature of the years-long timescale fading (or variability) observed by \citet{montet2016} and confirmed here in the NUV. Critically, with only a single wavelength band available and no apparent characteristic timescale for this variation with the 4-year observing window, little can be constrained from the \Kepler data alone. \citet{metzger2017} have argued the long-timescale fading could be due to stellar atmosphere recovery after a planetary in-spiral, and possibly the short-timescale dips are due to remaining debris. \citet{montet2016} note the fading in the \Kepler FFI's may be due to the transit of a dust cloud. However none of these models definitively explain the long-timescale variability observed in \citet{montet2016}.

By combining the optical \Kepler FFI light curve with the long-timescale GALEX NUV data presented here, we can place the first multi-wavelength constraints on KIC 8462852. A natural model to compare the simultaneous variability in the NUV and optical is that of a dust cloud. Extinction by dust in the interstellar medium is well studied, and several models with varying dust compositions are available at these wavelengths. Regardless of {\it where} the dust originates (i.e. circumstellar versus interstellar), such extinction models are a useful path forward in exploring the fading of KIC 8462852.

\begin{figure}[!t]
\centering
\includegraphics[width=3.5in]{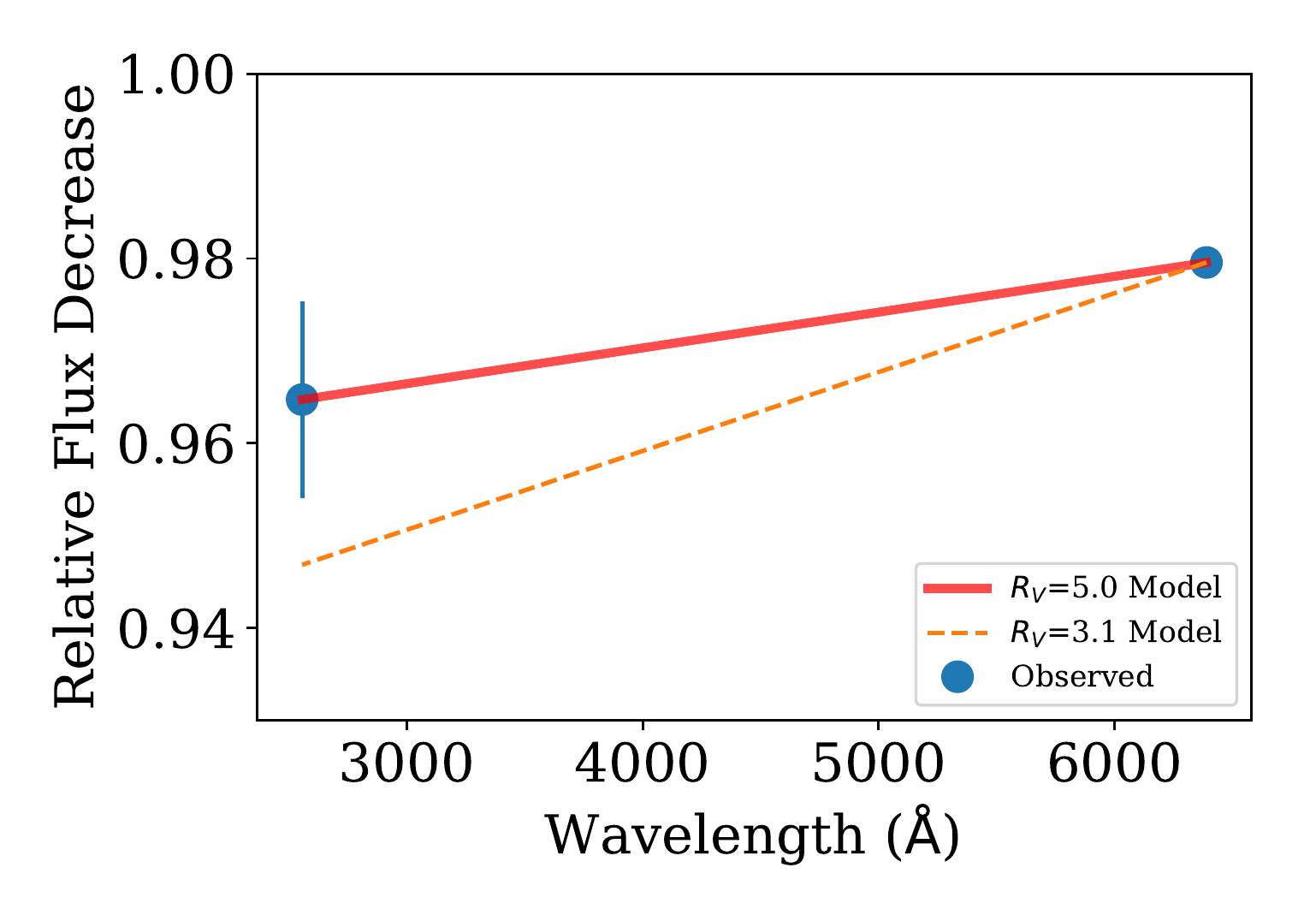}
\caption{Comparison between the flux decrease observed at the effective wavelengths of the GALEX NUV and \Kepler bands (blue circles), a corresponding $R_V=3.1$ dust model from \citet{cardelli1989} tuned to pass through the \Kepler data (orange dashed line), and a $R_V=5.0$ dust model that passes through both the \Kepler and NUV data (red solid line). The standard $R_V=3.1$ dust model over-predicts the NUV flux decrease given the observed \Kepler fading.}
\label{fig:dust}
\end{figure}

To demonstrate the impact dust would have in these two bands, we computed the extinction in the GALEX NUV band that would be predicted given the fading observed by \citet{montet2016} within the 2011 and 2012 time windows observed by GALEX. We used a standard \citet{cardelli1989} dust model with $R_V=3.1$, computed using the Python code from \citet{barbary2016}. The comparison of this $R_V=3.1$ prediction with the flux decrease observed by GALEX is shown in Figure \ref{fig:dust}. The \citet{cardelli1989} model over-predicts the fading found in the NUV, indicating the fading is more gray (less wavelength dependent) than a standard $R_V=3.1$ dust model. The NUV decrease measurement is 1.7$\sigma$ away from the  $R_V=3.1$ model, marginally inconsistent with ``normal'' interstellar dust as the culprit of the fading observed by \citet{montet2016},
and supports a circumstellar origin. Given the lack of warm circumstellar dust detection by \citet{thompson2016}, this material must be very cool.

However, the NUV response of dust models is highly dependent on grain composition. This can be explored in standard dust models by modifying the $R_V$ parameter. We then tuned a dust model to match both the observed \Kepler optical and GALEX NUV dimming by varying the $R_V$ and specific extinction ($A_V$) parameters. To fit the fading in both wavelengths simultaneously requires a dust model with  $R_V=5.0\pm0.9$. While this is not typical for interstellar extinction material, such a high $R_V$ has been reported for example around young protostars \citep[e.g.][]{hecht1982}. Competing dust models can produce significantly different NUV extinctions. For example, by modeling the \Kepler and GALEX fading for KIC 8462852 shown in Figure \ref{fig:longtime} with a \citet{fitzpatrick2009} dust model, we find a best-fit parameter of $R_V=5.8\pm1.6$. 
\citet{simon2017} found the dimming is weaker at redder optical wavelengths, broadly consistent with either dust extinction or temperature variations. A similarly large value for the reddening law ($R_V>5$) was recently reported for KIC 8462852 over a comparable timespan after the \Kepler mission using follow-up near-ultraviolet, optical, and NIR monitoring by \citet{meng2017}. However, the long timescale variation in $uvm2$-band flux was comparable to the intrinsic light curve uncertainty in their comparison stars. A joint analysis of the GALEX data presented here and the Swift/UVOT data from \citet{meng2017} may provide an improved understanding of the extinction properties of circumstellar dust around KIC 8462852.

Besides dust extinction, another simple model that can be invoked to fit the long timescale variations of KIC 8462852 is changes in the star's effective surface temperature. We carried out a toy model calculation of this cooling, assuming a quiescent blackbody temperature of 6750 K for the star, and flux variations in each wavelength band due to changes in blackbody temperatures. The \Kepler FFI fading seen by \citet{montet2016} over the same time windows as our GALEX observations requires a temperature change of 41$\pm$3 K. This temperature change in turn predicts a drop in the NUV flux of 5\%, which is in weak tension with our observed flux change of $3.5\pm1.0$\%.

\section{Summary}
\label{sec:summary}
We have undertaken the first exploration of the NUV variability for KIC 8462852, using GALEX data on a range of timescales. No significant variability is found on 10-100 second timescales using NUV light curves produced with {\tt gPhoton}. Over four visits spanning 70 days in 2011, we also find no significant medium-term variability.

Comparing co-added data from 2011 with the follow-up GCK study of the \Kepler field in 2012, we find that KIC 8462852 faded by $3.5\pm1.0$\% in the NUV. This fading coincides with the slow variation reported by \citet{montet2016}, and is the first verification that this star is variable in the NUV. A preliminary examination of the typical variance between the GALEX and GCK data shows an average NUV change of $\sim$3.5\% for bright F-type stars. Thus we believe the NUV fading observed for KIC 8462852 is real, but not necessarily atypical at these wavelengths. 

Though the long timescale NUV light curve is very sparsely sampled, the combination of NUV and optical wavelengths provides a powerful constraint on the nature of this slow dimming. We explored both dust extinction and thermal variations as possible causes for the long timescale fading. Our favored explanation from these NUV data is that KIC 8462852 may be occulted by a slowly changing column density of dust with $R_V=5$.

Finally, GALEX provides us with a valuable new dataset for use in the search for other objects of this class. We are able to expand the search criteria beyond dramatic short timescale events and slow dimming as observed with \Kepler, to now include slow variability in the NUV. If other F-type stars are found with similar multi-wavelength variability over long timescales, it will shed light important on the occurrence rate and possible lifetime of `Boyajian's Star'' type variables.

\acknowledgments
The authors wish to thank the anonymous referee, whose comments and suggestions improved the quality of this manuscript. 
We thank Joshua Simon for sharing a draft of their paper prior to publication.

JRAD is supported by an NSF Astronomy and Astrophysics Postdoctoral Fellowship under award AST-1501418. 

Work by B.T.M. was performed under contract with the Jet Propulsion Laboratory (JPL) funded by NASA through the Sagan Fellowship Program executed by the NASA Exoplanet Science Institute.

DJW has received funding from the European Research Council under the European UnionÕs Seventh Framework Programme (FP/2007-2013)/ERC Grant Agreement n. 320964 (WDTracer). 

Part of this research was carried out at the Jet Propulsion Laboratory, California Institute of Technology. under a contract with the National Aeronautics and Space Administration. This document does not contain export controlled information.

\software{gPhoton \citep[v1.28.2;][]{gphoton}, gatspy \citep{gatspy}, f3 \citep{montet2017}}


\end{document}